\documentclass[reprint,amsmath,amssymb,prl]{revtex4-2}
\usepackage[pdfstartview=FitH,
            breaklinks=true,
            colorlinks=true,
            linkcolor=black,
            citecolor=black,
            urlcolor=black,
            pdftitle={Fluctuating environments are sufficient to drive substantial variability in species
abundance across locations},
            pdfauthor={James Henderson},
            ]{hyperref}
\usepackage[inline]{enumitem}
\usepackage{comment}
\usepackage{amsmath}
\usepackage{amsfonts}
\usepackage{amssymb}
\usepackage{mathtools}
\usepackage{graphicx}
\makeatletter
\newcommand{\NoTOC}{\let\addcontentsline\@gobblethree}
\makeatother
\usepackage{bibunits}
\defaultbibliographystyle{apsrev4-2}
\defaultbibliography{bibliography}

\newcommand{\mfcorr}{\zeta} 

\begin{document}

\begin{bibunit}
\title{Fluctuating environments are sufficient to drive substantial variability in species abundance across locations}

\author{James F. D. Henderson}
\affiliation{Division of Infection and Immunity \& Institute for the Physics of Living Systems, University College London}
\author{Andreas Tiffeau-Mayer}
\affiliation{Division of Infection and Immunity \& Institute for the Physics of Living Systems, University College London}
\date{\today}

\begin{abstract}
 Species growing in environments that change in time and space will vary in their abundance across locations, even in the absence of persistent location preferences. Here we quantify this non-equilibrium effect by studying a minimal model of a spatially compartmentalised community with time-averaged-neutral competition but location-dependent environmental fluctuations. We analytically derive distributions of two-point inequality, defined as the log-ratio of a species' abundance across a pair of locations. We characterise how the balance of relaxation via migration and fluctuation strength determine the bulk and extreme value statistics of these distributions in the two-patch and infinite-patch cases. We demonstrate the existence of a noise-induced transition to bimodal inequality, which depends on the correlation timescale of the environmental fluctuations.  Finally, we discuss the evolutionary benefit of finite migration rates in environments with temporal correlations.
\end{abstract} 

\maketitle

That species fluctuate in their abundance across space and time is an ecological rule, not an exception \cite{rosenzweig1996species}, however identifying the key drivers of these fluctuations in high-dimensional ecosystems remains an open challenge~\cite{ grilli_macroecological_2020, rogers2022chaos}. Changing environmental conditions act as one source of such fluctuations, as they alter the  availability of resources required for a species to prosper and reproduce \cite{levin1976population, ottino2020population, asker2025fixation}. Additionally, interactions between species in large communities can drive dramatic abundance changes and chaotic dynamics, acting as an effective stochastic driving on single-species growth rates \cite{pearce2020stabilization, de2025self}. Previous work has thus considered how the interplay of spatial compartmentalisation and random growth affects extinction times \cite{agranov2021extinctions} and has recently highlighted the ability of noise and space to fundamentally alter coexistence regimes~\cite{al2026spatiotemporal}. Here, we build on this work by focusing on the basic question at the core of this phenomenology: How much spatial variability in a species' abundance should we expect from environmental fluctuations alone, in an otherwise neutral community? 

Metacommunity theory provides a framework to describe spatially compartmentalised communities consisting of $S$ species migrating between $N$ locations (patches)~\cite{leibold2004metacommunity, garcia2024interactions}. In the general case, the population size, $C_{k, i}$, of species $k$ in patch $i$ evolves according to
\begin{equation}
\begin{split}
\label{population equation}
    \frac{dC_{k,i}}{dt} = &\left[r_{k,i}\left(\vec{C}_i\right) + \eta_{k,i}(t)\right]C_{k,i}  \\ &+ \sum_{j \neq i}^{N-1} \left(M_{k,ij} C_{k,j} - M_{k,ji}C_{k,i}\right),
\end{split}
\end{equation}
for $i = 1, ...,N$ and $k = 1, ...,S$. The terms $r_{k,i}(\vec{C}_i)$ account for interactions between species. Each $\eta_{k,i}(t)$ models the impact of temporal fluctuations in the environmental conditions within patch $i$ on the growth rate of species $k$. For simplicity, we will neglect demographic noise which would introduce sub-leading terms $\sim\sqrt{C_{k,i}}\xi_{k,i}(t)$. $M_{k,ij}$ defines the rate of migration of individuals of species $k$ into patch $i$ from patch $j$.

This framework allows us to specify alternate explanations for spatial heterogeneity in species abundance: A species can vary in its abundance between locations because of persistent niche differences (modelled by $r_{k,i}(\vec{C_i})$), migratory preferences ($M_{k, ij}$), or alternatively as a non-equilibrium effect of transient fitness fluctuations ($\eta_{k,i}(t)$). We are interested here in this latter setting, in which species have no inherent preference for a particular location. We therefore assume symmetric migration with $M_{k, ij} = M / N$, where $M$ is the overall coupling strength between patches, and consider time-averaged neutral competition where all species compete uniformly and are equally fit on average across locations~\cite{mallmin2025fluctuating}. In the limit $S \rightarrow  \infty$, such competition results in the decoupling of species dynamics, with $r_{k, i}(\vec{C}_i) \rightarrow r \leq 0$, allowing for the species index $k$ to be dropped~(see~SI). Eq.~\eqref{population equation} effectively reduces to the dynamics of a single species migrating between multiple locations (a metapopulation \cite{levins1969some, hanski1998metapopulation, evans2013stochastic}).

We expect this setting to provide a useful null model for the study of spatial compartmentalisation in large ecological communities, as it represents the simplest extension of the time-averaged neutral theory of ecology~\cite{volkov2003neutral, desponds2016fluctuating,gaimann2020early,kessler2024dominance, mallmin2025fluctuating, azaele2016statistical, hubbell2011unified} to the spatial domain. For instance, in the adaptive immune system populations of lymphocytes circulate the body and undergo periods of rapid growth upon antigen encounter. Time-averaged neutral theory has previously been used to explain heavy-tailed species abundance distributions in this system \cite{desponds2016fluctuating, gaimann2020early}. However, it is unknown whether local expansion and migration might suffice to also explain observed variation in clonal lineage (species) abundance of immune cells across locations \cite{dewolf2023tissue, sureshchandra2025deep}. Beyond the field of ecology, this setting provides a stylised model for how market fluctuations and wealth exchange drive inequality between economic agents \cite{bouchaud_wealth_2000, bouchaud2015growth}, and also relates to the KPZ equation of random surface growth~\cite{kardar1986dynamic}.

Even in this simple setting, we will find a surprisingly rich phenomenology, depending on the dimensionality of the system $N$ and the statistical structure of the noise $\eta_{k,i}(t)$.  We restrict our study to spatially uncorrelated  environmental fluctuations modelled as stochastic processes of zero mean and intensity of $2D$, defined as the zero-frequency part of their stationary power spectra. Previous work studying a non-spatially compartmentalised species has shown that the correlation timescale of environmental noise, i.e. the colour of the noise, can significantly alter the risk of species extinction~\cite{spanio2017impact, kamenev2008colored}. We therefore allow the noise to be temporally correlated and study how emerging species fluctuations depend on the environmental correlation time $1/\lambda$.  

In the study of this system, the two-point inequality of species abundances
\begin{equation}
x_{ij} = \ln\left(\frac{C_i}{C_j} \right),
\end{equation}
provides a natural observable characterising the fluctuations in a single species' abundance between a pair of locations. In the following we thus present analytical solutions for distributions of $x_{ij}$ in the $N=2$ and $N\rightarrow\infty$ cases for white and coloured environmental noise.

\begin{figure}[h!]
\includegraphics[width=1\columnwidth]{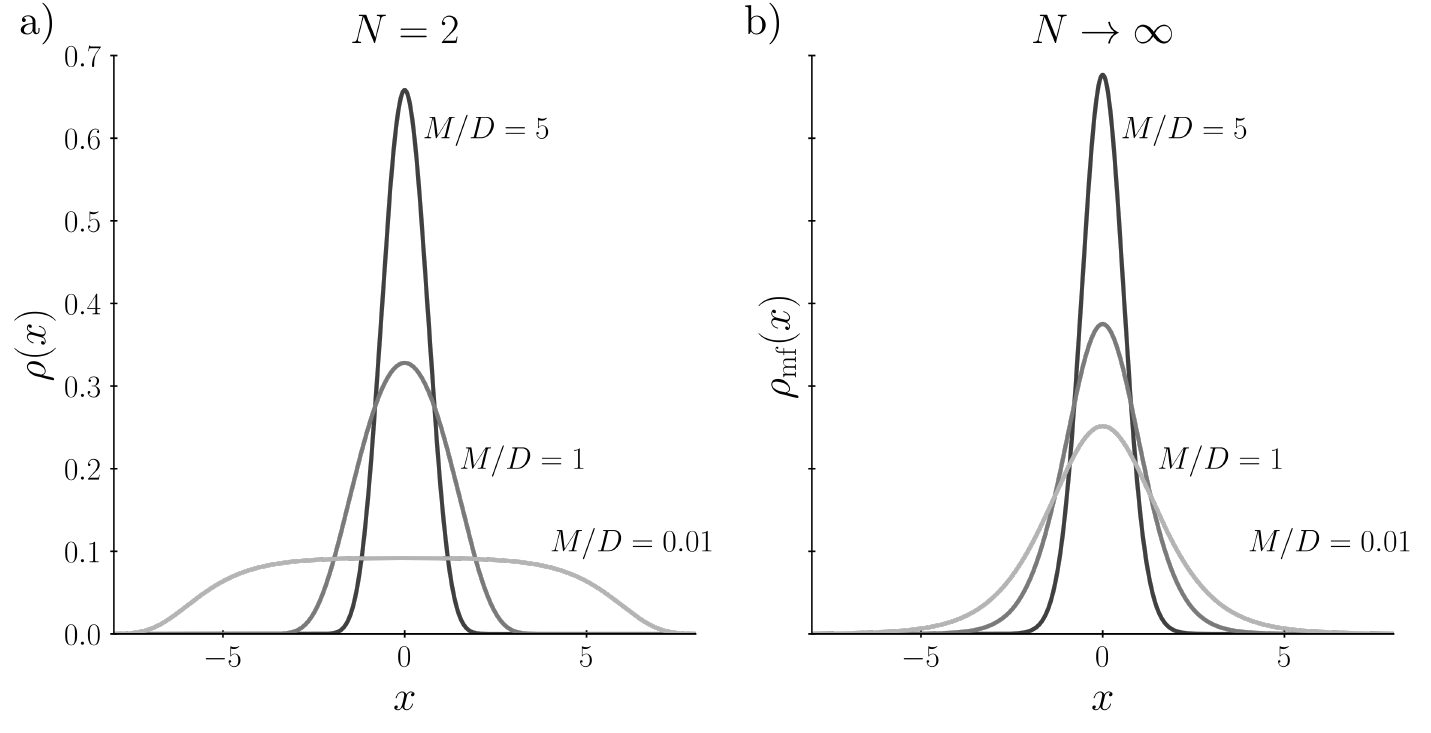}
\caption{Analytical densities for the steady-state distribution of two-point inequality in the case of a) $N=2$  given by Eq.~\eqref{white noise steady state x} and b) $N \rightarrow \infty$ given by Eq.~\eqref{white noise steady state x mf} for a range of migration rate to noise intensity ratios.}
\label{white_noise_densities}
\end{figure}

\emph{Temporally uncorrelated environments--} The simplest scenario in which this system can be studied considers temporally uncorrelated environmental fluctuations, $\left<\eta_i(t)\eta_i(t^{\prime})\right> =  2D_i\delta(t - t^{\prime})$. For $N=2$, working in the Stratonovich convention, the log-ratio $x = x_{12}$ evolves autonomously according to \cite{agranov2021extinctions}
\begin{equation}
\label{sde for log ratio}
    \frac{dx}{dt} = -M \sinh x +\eta_x(t),
\end{equation}
where $\eta_x(t) = \eta_1(t) - \eta_2(t)$ and has intensity $4D$. To obtain the steady state distribution of $x$, we recast Eq.~\eqref{sde for log ratio} into its corresponding Fokker–Planck equation \cite{gardiner2009stochastic}
\begin{equation}
    \partial_t \rho(x, t) = -\partial_x\left[-M\sinh (x) \rho(x, t) - 2D \partial_x \rho(x,t)\right].
\end{equation}
Solving for $\partial_t \rho(x, t) = 0$ with zero flux boundary conditions shows that the system approaches a non-equilibrium steady-state, where two-point inequality is distributed according to \cite{agranov2021extinctions, bouchaud2015growth}
\begin{equation}
\label{white noise steady state x}
    \rho(x) = \rho(x, \infty) = \frac{1}{2\mathcal{K}\left(0, \frac{M}{2D}\right)}e^{- \frac{M}{2D}\cosh x},
\end{equation}
where $\mathcal{K}(\nu, z)$, is a modified Bessel function of the second kind. Around the origin, this distribution is approximately Gaussian, with a double exponential suppression of large values of $|x|$ with asymptotic scaling $\rho(x) \sim e^{-M/(4D) e^{|x|}}$. In the limit of weak coupling, $M / D \ll 1$, this distribution is approximately uniform for small values of $|x|$ (Fig.~\ref{white_noise_densities}a), indicating that many intermediate values of two-point inequality are similarly likely. 

The white-noise driven system may also be studied in the limit of $N\rightarrow\infty$ by making use of a mean-field approximation (see SI). In such an approximation each site evolves independently and $y_i = \ln(C_i / \bar{C})$, where $\bar{C} = \frac{1}{N} \sum_{i}C_i$ is the average population size across all patches, approaches a steady-state density. This mean-field steady-state density is \cite{bouchaud_wealth_2000, ottino2020population}
\begin{equation}
\label{white noise steady state y mf}
    \rho_{\mathrm{mf}}(y) = \frac{\left(\frac{M}{D}\right)^{(\frac{M}{D} +1)}}{\Gamma\left(\frac{M}{D}+1\right)} e^{-\frac{M}{D} e^{-y} - (\frac{M}{D}+1)y},
\end{equation}
where we have dropped the site-specific subscript. At large $y$, we recover the well-known power law scaling of $w = e^y$  occurring in convergent multiplicative processes repelled from zero \cite{sornette1997convergent}. Using $\rho_{\mathrm{mf}}(y)$, the density for $x_{ij} = y_i - y_j, i \neq j$, may be obtained using the fact that $y_i, y_j \overset{i.i.d}{\sim} \rho_{\mathrm{mf}}(y)$ and hence
\begin{align}
\label{convolution}
    \rho_{\mathrm{mf}}(x) &= \int_{-\infty}^{\infty} \rho_{\mathrm{mf}}(y)\rho_{\mathrm{mf}}(y+x) dy \\
    \label{white noise steady state x mf}
    &= \frac{\Gamma\left(2\left(\frac{M}{D}+1\right)\right)}{\Gamma\left(\frac{M}{D}+1\right)^2}\left(1+e^{-x}\right)^{-2\left(\frac{M} {D}+1\right)} e^{-(\frac{M}{D} +1)x},
\end{align}
where $\Gamma(z)$ is the Gamma function. 

In contrast to the two patch case, this distribution has heavier tails which scale $\rho_{\mathrm{mf}}(x) \sim e^{-(M/D + 1)|x|}$, and approaches a limiting density for weak coupling (Fig.~\ref{white_noise_densities}b and see SI). Taken together, these results reveal an interesting distinction for two-point inequality between the low and high dimensional regimes whose distributions differ in both bulk and tail behaviour.

\emph{Temporally correlated environments--} As most real sources of environment variability will exhibit temporal correlations, we next turn to the study of Eq.~\eqref{population equation} when driven by coloured noise with correlation timescale $1/\lambda$. We adopt the simplest model for such temporal correlations, an Ornstein-Uhlenbeck process defined by
\begin{equation}
\label{sde for eta}
    \frac{d\eta_i}{dt} = -\lambda \eta_i(t) + \lambda\sqrt{2 D}\xi_i(t),
\end{equation}
where each $\xi_i(t)$ is a zero-mean Gaussian white noise.

\begin{figure*}
\begin{center}
\includegraphics[scale=0.7]{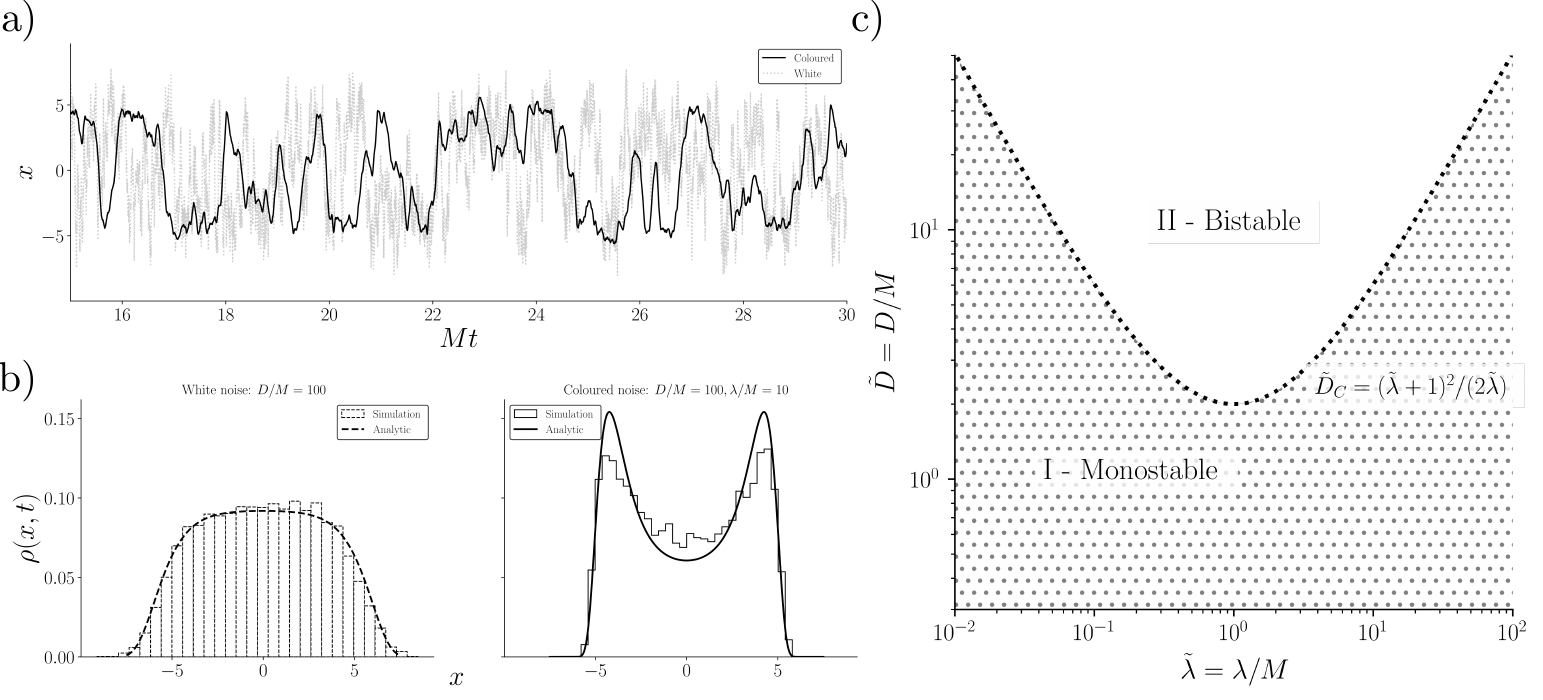}
\caption{a) Trajectories of Eq.~\eqref{sde for log ratio} simulated via the Euler-Maruyama method for white environmental noise (dotted line) with $M = 0.1$, $D=10$, and a coloured noise (solid line) with $M = 0.1$, $D=10$ and $\lambda = 1$. b) Histograms of trajectories obtained at $Mt = 20 $ over $2 \times 10^4$ simulation runs. The analytic steady state densities are overlaid. Simulation runs are initialised as $x(0)=0, \eta(0)=0$. c) `Phase' diagram highlighting monostable and bistable parameter regimes. These regions are separated by a supercritical pitchfork bifurcation \cite{strogatz2024nonlinear}  shown by the dashed black line.}
\label{phase_diagram}
\end{center}
\end{figure*}

For certain parameter values with $N=2$, simulations reveal that $x(t)$ unexpectedly switches between two typical values of two-point inequality, in contrast to simulations with white noise at equal noise intensity (Fig.~\ref{phase_diagram}a). In line with this, histograms of $x$ show that its steady state distribution is bimodal when driven by coloured noise, as opposed to the unimodal distribution observed for white noise (Fig.~\ref{phase_diagram}b). The rare but rapid switches in the bimodal case are reminiscent of relaxation oscillations in non-linear oscillators such as the Van der Pol system \cite{strogatz2024nonlinear}.
These numerical results are suggestive of a noise-induced transition, a class of phenomena in which a system driven by noise can enter states not present in its deterministic dynamics \cite{horsthemke1984noise, coomer2022noise}.

To analytically characterise this noise-induced transition, we made use of the Unified Coloured Noise Approximation \cite{jung1987dynamical} (see SI) to obtain an effective 1D stochastic differential equation for $x$
\begin{equation}
\label{ucna sde}
    \frac{dx}{dt} = \frac{-M\sinh x}{1+\frac{M}{\lambda} \cosh x} + \frac{\sqrt{4D}}{1+\frac{M}{\lambda}\cosh x} \xi_x(t),
\end{equation}
where $\xi_x(t)$ is a zero-mean Gaussian white noise taken with the Stratonovich prescription. As before, Eq.~\eqref{ucna sde} may be converted into a Fokker–Planck equation whose steady state density can be written in terms of an effective potential $U_{\mathrm{eff}}(x)$
\begin{equation}
\label{coloured noise steady state x}
    \rho(x) = \frac{1}{Z} e^{-U_{\mathrm{eff}}(x)},
\end{equation}
where $Z$ is a normalisation constant, and
\begin{align}
\label{coloured_noise_phi}
U_{\mathrm{eff}}\left(x\right)
 = &\frac{M}{2D}\left(1 + \frac{M}{2\lambda} \cosh x\right) \cosh x \nonumber \\ &- \ln \left(1 + \frac{M}{\lambda } \cosh x\right).
\end{align}
Analysis of the extrema of this effective potential allows us to derive a phase diagram for the bistable region of parameter space (Fig.~\ref{phase_diagram}c).

The transition happens above a critical noise intensity $\tilde{D} > \tilde{D}_c = (\tilde{\lambda} + 1)^2/(2\tilde{\lambda})$, which depends on the non-dimensional ratios $\tilde{D} =D/M$ and $\tilde{\lambda} = \lambda/M$. Above this critical value the effective potential has a single maxima at the origin and two minima located at $x^{*}_{\pm} = \pm\cosh^{-1}\left(\sqrt{2\tilde{D}\tilde{\lambda}} - \tilde{\lambda}\right)$ implying a bimodal steady-state density. The predictions by the unified coloured noise approximation agree closely with numerical simulations (see SI). In the limit of fast environmental decorrelation with respect to the rate of migration, $\tilde{\lambda} \rightarrow \infty$, we fully recover the white noise density as expected. In the opposite, `quenched', limit of a highly persistent environment, $\lambda \rightarrow 0$ at finite $D\lambda = \sigma^2$, we find that the transition occurs at $\sigma^2 \geq M^2 / 2$. This is directly equivalent to an ensemble of two-patch systems with disordered basal growth rates, $r_i$, drawn from a zero-mean Gaussian distribution of variance $\sigma^2$ (see SI).

We now turn to $N \rightarrow \infty $ for a temporally correlated environment. The dynamics of $y_i = \ln(C_i / \bar{C})$ are (see SI)
\begin{equation}
\label{eqn for y}
    \frac{dy_i}{dt} = M(e^{-y_i} -1) + \eta_i(t) - \mfcorr(t)
\end{equation}
where $\mfcorr(t) = \frac{1}{N}\sum_{i} \frac{C_i(t)}{\bar{C}(t)} \eta_i(t)$. We assume self-averaging in the mean-field limit, $\mfcorr(t) \rightarrow \mfcorr$, which can be solved for by the self-consistency condition $\left<C_i\right>/\bar{C} = 1$. We once again employ the Unified Coloured Noise Approximation and find the steady state density for $y$ to be $\rho(y) \propto \exp(-U_{\mathrm{eff, mf}}(y))$, where
\begin{widetext}
\begin{equation}
\label{coloured noise steady state y mf}
U_{\mathrm{eff, mf}}(y) =  \frac{1}{D}(M + \mfcorr)y  +\frac{M}{D}\left(1 - \frac{1}{\lambda}(M + \mfcorr)\right) e^{-y}+\frac{M^2}{2\lambda D}e^{-2y} - \ln \left(1 + \frac{M}{\lambda}e^{-y}\right). 
\end{equation}
\end{widetext}
Large positive fluctuations in $y$ are controlled by the first term, which again implies power-law scaling for $w= e^y$ for $w \gg 1$. Following Eq.~\eqref{convolution} to derive the steady-state density for $x$, we find that it inherits the large $y$ scaling, with $\rho_{\mathrm{mf}}(x) \sim e^{-|x|(M+\zeta)/D } $~(see SI). 

As $\zeta$ defines the tail behaviour of the distributions, it dictates the extreme value of statistics of the system and controls the divergence of distribution moments.
For any finite model parameters, the self-consistency condition holds, implying a finite mean for $w$ (see SI). However, in the limit of `quenched' disorder, $\lambda \rightarrow 0$, self-consistent solutions only exist for $D < M$ above which the mean-field approximation breaks down~(see SI).  Motivated by \cite{bernard2025mean}, which studied disordered $r_i$'s in the limit of large $N$, we find that when scaling parameters as $D\lambda = \sigma^2 / (2\ln N)$ while keeping $2 \lambda\ln N  = M$ the breakdown of the mean-field approximation occurs at $\sigma^2 \geq M^2 $. This corresponds to the so-called `localisation' transition at which a single patch holds a finite fraction of the system's biomass.

\begin{figure}[h!]
\includegraphics[width=\columnwidth]{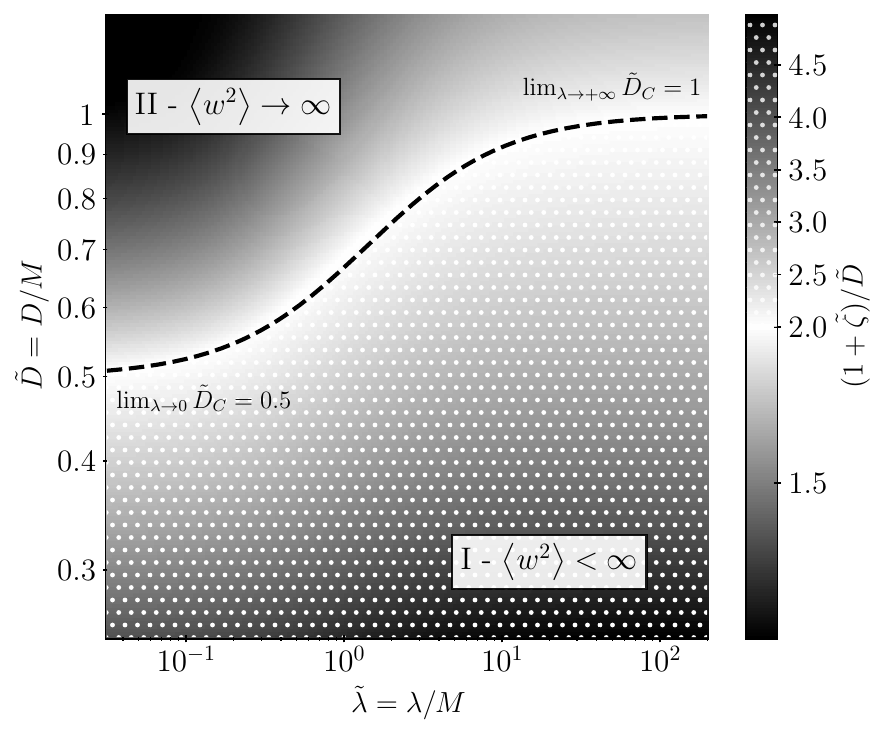}
\caption{Tail exponent of two-point inequality in the mean-field limit obtained by numerically solving the self-consistency equation $\left<e^{y}\right> = \left<w\right>=1$. The dashed line is the boundary predicted by a saddle-point approximation (see text). }
\label{tail exponent}
\end{figure}

A second transition, referred to as the `condensation' transition \cite{ichinomiya2012bouchaud}, occurs when $M  + \zeta < 2 D$ which causes $\left<w^2\right>$ to diverge. For a white driving noise $\zeta = D$ and so $\left<w^2\right>$ diverges for $D > M$ \cite{ottino2020population}. Solving for $\zeta$ numerically we find that for coloured noise this transition can occur at a lower noise intensity~(Fig.~\ref{tail exponent}). In the weak-noise limit, $D < M$, a saddlepoint approximation allows us to show that
\begin{equation}
    \zeta \approx D \left(\frac{\lambda}{M + \lambda}\right) + \mathcal{O}\left(\left(\frac{D}{M}\right)^2\right).
\end{equation}
Accordingly, the transition to a `condensed' phase is predicted to occur for $\tilde{D} \geq \tilde{D}_c=  1/(2 - \tilde{\lambda} / (1  +\tilde{\lambda}))$, which is in good agreement with the numerical results~(Fig.~\ref{tail exponent}).

\emph{Optimal dispersal in time varying environments--} In addition to controlling the divergence of moments, $\zeta$ also sets the logarithmic growth rate of the population
\begin{equation}
\left<\frac{d\ln\bar{C}}{dt}\right> = r + \left<\zeta(t)\right>.
\end{equation}
 Our results thus far therefore relate to the stability and long-term evolutionary success of populations living in time-varying environments \cite{kelly1956new, rivoire2011value,al2026spatiotemporal}. We can naturally ask which rate of migration $M$ optimises long-term growth in a fluctuating environment with fixed $\lambda$ and $D$. For $N \rightarrow \infty$ and finite $\lambda$, we find that $\zeta$ monotonically decreases with $M$ and hence the optimal strategy is to not migrate (see SI). Intuitively, in an infinite system we might always expect some patch to experience favourable environmental conditions that persist for some finite time and dispersal then merely acts to dilute this growth across space.

\begin{figure}[h!]
\includegraphics[width=0.7\columnwidth]{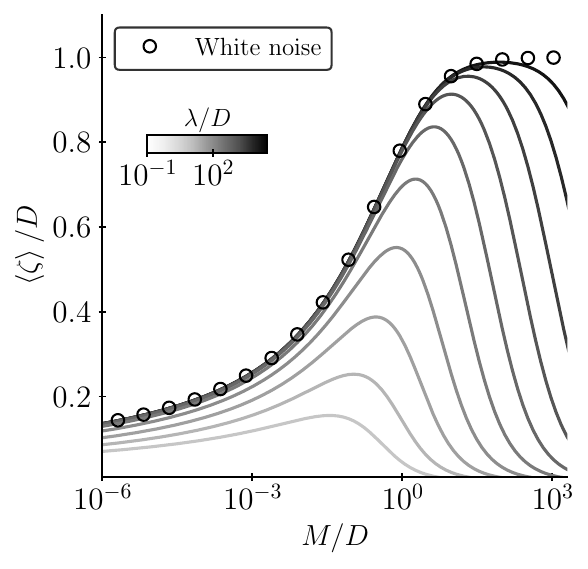}
\caption{Logarithmic growth rate for a species migrating between two time varying environments as a function of its migration rate. Each curve shows the numerically obtained growth rates for fluctuations with different correlation timescales $1/\lambda$. Circles show the analytic solution for temporally uncorrelated environmental fluctuations.  All rates have been rescaled by the intensity of the environmental fluctuations, $D$, and $r$ set to $0$.}
\label{optimal_growth}
\end{figure}

In stark contrast, numerical evaluation of $\zeta$ for $N =2$ shows that the growth rate peaks at a finite value of $M$ which scales with the degree of environmental persistence $\lambda$ (Fig.~\ref{optimal_growth}). We expect such an optimal migration rate to exist for any finite $N$ and to emerge as a balance of exploiting temporally-correlated streaks of good luck while rebalancing the growing population to hedge against bad times. In the context of economics and finance, where $\zeta$ is related to the geometric mean return, this result implies the existence of an optimal rate of portfolio rebalancing or wealth-taxation \cite{bouchaud2015growth} to exploit the empirically well-documented temporal correlations in real market returns~\cite{jegadeesh1993returns}.

\emph{Conclusion--} We have provided a unified theory describing the extent to which environmental fluctuations drive spatial species variability in a simple time-averaged neutral setting. Our results extend previous work~\cite{ottino2020population,agranov2021extinctions,evans2013stochastic,bouchaud_wealth_2000} to show how the bulk and extreme values of spatial inequality are controlled by the interplay of system timescales and dimensionality. We find significant differences in the distribution of inequality for a small ($N=2$) and large ($N \rightarrow \infty$) number of patches, with low dimensions allowing for broader bulk behaviour at weak couplings ($D < M$) while exhibiting super-exponential suppression of extreme values of inequality. We expect this super-exponential suppression to arise for extreme value of $|x|$ at any finite $N$, due to one patch becoming dominant and driving fluctuations in the total population size~\cite{medo2009breakdown, liu2018absence}. Temporal correlations in environmental fluctuations significantly change system behaviour, and lead to a coloured noise induced transition in the low dimensional system. In high dimensions, temporal correlations produce heavier tailed distributions of inequality and lower the noise strength for which the second moment of population sizes diverges. In the limit of quenched random growth, extremely broad distributions with a diverging first moment are produced as recently discussed by Ref.~\cite{bernard2025mean}. Finally, we demonstrate the evolutionary benefit of a hot-hand strategy in correlated environments, where finite migration rates enable populations to exploit localised streaks of beneficial environmental conditions. This effect has been previously described in the context of diversification strategies in time-varying environments with correlations \cite{ratcliff2015courting}, but stands in contrast to classic bet-hedging strategies in temporally uncorrelated environments, where instant rebalancing of population abundances is optimal \cite{kelly1956new}.

The distributions identified in this work can provide null models for the study of spatially distributed communities in fluctuating environments. In this sense, we plan to apply our results to the adaptive immune system, to quantify the extent to which variations in clonal abundance across bodily sites can be explained by localised antigen fluctuations. This system also motivates extending the model to include disordered migration rates as an alternative mechanism for spatial variability, as it is known that populations of immune cells adopt varying migration strategies given their phenotypic state \cite{masopust2013integration, krummel2016t, christo2024multifaceted}. Given the large differences in system behaviour across parameter regimes, future empirical work on the inference of migration rates and environmental fluctuation statistics would be particularly valuable in driving model selection. 
{\NoTOC
\begin{acknowledgments}\textbf{Acknowledgments} The work of J.F.D.H and A.T.-M. was supported by the National Institute for Health and Care Research University College London Hospitals Biomedical Research Centre. The authors acknowledge helpful discussions with Reemon Spector, Antonio Matas Gil and Guy Bunin.
\end{acknowledgments}
}

{\NoTOC
\putbib
}

\end{bibunit}

\newpage

\onecolumngrid
\begin{bibunit}
\setcounter{equation}{0}
\setcounter{figure}{0}
\renewcommand{\thefigure}{S\arabic{figure}}
\setcounter{secnumdepth}{3}
\begin{center}
	\textbf{\large  Fluctuating environments are sufficient to drive substantial variability in species abundance across locations
		\\
		[.3cm] -- Supplemental Material --} \\
	[.4cm] James F. D. Henderson and Andreas Tiffeau-Mayer \\ [.1cm]
	{\itshape Division of Infection and Immunity \& Institute for the Physics of Living Systems, University College London}\\
\end{center}

\setcounter{tocdepth}{3}
\tableofcontents

\section{The Unified Coloured Noise Approximation (UCNA)}

To make our derivations self-contained, we briefly review the Unified Coloured Noise Approximation \cite{jung1987dynamical}. Consider the following system driven by an Ornstein-Uhlenbeck process with relaxation time $1/\lambda$ and intensity $2D$,
\begin{align}
\label{sde for x}
\frac{dx}{dt} &= F(x)+\eta(t), \\
\label{sde for eta}
  \frac{d\eta}{dt} &= -\lambda \eta(t) + \lambda\sqrt{2 D}\xi(t),
\end{align}
where $\xi(t)$ is a zero-mean, unit-strength, Stratonovich white noise.
We take the time derivative of Eq.~\eqref{sde for x} and substitute Eq.~\eqref{sde for eta} 
to obtain a second order differential equation for $x$
\begin{equation}
\begin{aligned}
    \frac{d^2x}{dt^2} &= F^\prime(x) \frac{dx}{dt} + \frac{d\eta}{dt}  \\
    &= F^\prime(x) \frac{dx}{dt} -\lambda \eta(t)  + \lambda\sqrt{2D}\xi(t),
\end{aligned}
\end{equation}
Where $F^\prime(x)$ denotes differentiation with respect to $x$. We now use Eq.~\eqref{sde for x} to eliminate $\eta$
\begin{equation}
    \frac{d^2x}{dt^2} +\frac{dx}{dt}\left(\lambda - F^\prime(x)\right) =  \lambda F(x)   + \lambda\sqrt{2D}\xi(t)
\end{equation}
Following the UCNA \cite{jung1987dynamical}, we introduce a new timescale $u = t \sqrt{\lambda}$, which gives 
\begin{equation}
    \frac{d^2x}{du^2} +\gamma(x, \lambda)\frac{dx}{du} = F(x) + \sqrt{2D}\xi(u/\sqrt{\lambda})
\end{equation}
where we have introduced the non-linear `damping' factor
\begin{equation}
    \gamma(x, \lambda) = \sqrt{\lambda}\left(1-\frac{1}{\lambda}F^\prime(x) \right)
\end{equation}
in the regime when $\gamma \gg1$, we set $d^2x/du^2 = 0$. This holds for $\lambda \gg 1$ and $\lambda \ll 1$ (with the validity of the small $\lambda$ limit depending precisely on $F(x)$ \cite{spanio2017impact}). We obtain
\begin{equation}
    \frac{dx}{du} =  \frac{F(x) }{\gamma(x, \lambda)}  + \frac{\sqrt{2D}}{\gamma(x, \lambda)}\xi(u/\sqrt{\lambda}).
\end{equation}
Shifting back to $t=u/\sqrt{\lambda}$, the dynamics are now
\begin{equation}
    \frac{dx}{dt} =  \frac{F(x)}{1-\frac{1}{\lambda} F^\prime(x)}  + \frac{\sqrt{2D}}{1-\frac{1}{\lambda} F^\prime(x)}\xi(t).
\end{equation}
In the main text we use this general equation to derive effective state-dependent stochastic drift and diffusion coefficients, which in turn imply effective potentials in the Fokker-Planck description.

\section{Mean-field approximations}
\subsection{Metacommunity with time-averaged neutral competition}

We consider a community of $S$ species residing over $N$ spatial locations (a metacommunity). We take the population size of species $k$ in location $i$ to evolve according to
\begin{equation}
    \frac{dC_{k,i}}{dt} = \left[r_{k,i}\left(\vec{C}_i\right) + \eta_{k,i}(t)\right]C_{k,i} + \sum_{j \neq i}^{N-1} \left(M_{k,ij} C_{k,j} - M_{k,ji}C_{k, i}\right),
\end{equation}
for $i = 1, ...,N$ and $k = 1, ...,S$. Each $\eta_{k,i}(t)$ is a zero-mean noise modelling the influence of a fluctuating environment on the species' growth rate. Each $M_{k,ij}$ is the rate of migration of individuals of species $k$ into patch $i$ from patch $j$. We will consider the system to be homogeneously coupled and take all $M_{k,ij} = M/N$. The terms $r_{k,i}(\vec{C}_i)$ account for interactions between species. Here we consider a model of `neutral competition' \cite{mallmin2025fluctuating} in which species compete locally for a single shared resource with no species having an intrinsic advantage over any other. We will consider the following form of neutral competition
\begin{equation}
    \label{neutral competition}
    r_{k,i}\left(\vec{C}_i\right) = \frac{b}{T_i} - d,
\end{equation}
where $T_i = \sum_{k=1}^S C_{k,i}$ is the total number of individuals within patch $i$, and $b$ and $d$ are basal birth and death rates, respectively. This form of competition is motivated by our interest in the study of the adaptive immune system, with similar competition used to describe the population dynamics of T cells competing for cytokines \cite{gaimann2020early, de1994t}. Under such competition, $T_i$ evolves according to
\begin{align}
    \frac{dT_i}{dt} &= sC_0+\sum_{k=1}^{S} \frac{dC_{k,i}}{dt} \nonumber \\
    &= sC_0 + b - dT_i +\frac{M}{N}\sum_{j\neq i}^{N-1} (T_j - T_i) + \Xi(t),
\end{align}
where we have included an influx of new species arriving at a rate $s$ and with an initial population size per patch of $C_0$. This term can however be neglected in the case of a community with a fixed number of species. We have defined 
\begin{equation}
    \Xi(t) = \sum_{k=1}^S \eta_{k,i}(t) C_{k,i}(t).
\end{equation}
If we assume that the timescale at which the total population sizes per patch approach their steady state values is far longer than the correlation timescale of the environment driving random growth, we can take the $\eta_{k,i}$'s to be uncorrelated zero-mean white-noises in the Stratonovich convention each with intensity $2D$ such that $\left<\eta_{k,i}(t)\eta_{l,j}(t^\prime)\right> = 2D \delta_{kl}\delta_{ij}\delta(t - t^\prime)$. This simplification is supported by the analyses from Ref.~\cite{mallmin2025fluctuating} which show that for a fixed environmental noise variance ($D\lambda$), the correlation timescale only weakly impacts single patch community composition. To make progress, we convert the noise into the It\^o form
\begin{equation}
    \eta_{k,i}(t)C_{k,i} dt =  \sqrt{2D}C_{k,i}\circ dB_{k,i} = \sqrt{2D}C_{k,i} dB_{k,i}  + DC_{k,i} dt,
\end{equation}
and so 
\begin{align}
    \Xi(t) dt &= \sum_{k=1}^S\left(\sqrt{2D}C_{k,i} dB_{k,i}  + DC_{k,i} dt,\right)  \nonumber \\ &= DT_i dt + \sum_{k=1}^S \sqrt{2D}C_{k,i} dB_{k,i} \nonumber \\ 
    &= DT_{i}dt + \sqrt{2D\sum_{k=1}^S C_{k,i}^2(t)}dB_{i} 
\end{align}
where $dB_{i}$ is a new Brownian increment. Assuming no species holds a macroscopic fraction of the total biomass within the patch, then
\begin{equation}
\label{even approximation}
   \sqrt{\sum_{k=1}^S C_{k,i}^2}dB_{i} =  T_i  \sqrt{\sum_{k=1}^S \left(\frac{C_{k,i}}{T_i}\right)^2}dB_{i} \;\xrightarrow[S\to\infty]{}\; 0.
\end{equation}
Under such an approximation, the total population size in a given patch evolves as 
\begin{equation}
     \frac{dT_i}{dt} \approx sC_0 + b +(D - d)T_i +\frac{M}{N}\sum_{j\neq i}^{N-1} (T_j - T_i) .
\end{equation}
We assume that there exists a steady state for the system such that $dT_i/dt |_{T_i = T_i\star} = 0$ for all $i$ and hence
\begin{equation}
\label{fixed point eqn}
     sC_0 + b +(D - d)T^{\star}_i +\frac{M}{N}\sum_{j\neq i}^{N-1} (T^{\star}_j - T^{\star}_i) = 0,
\end{equation}
where each $T_i^{\star}$ is the steady state value of $T_i$. We now define $P^{\star} = \sum_{i=1}^N T_{i}^\star$ to be the total number of individuals in the entire system such that Eq.~\eqref{fixed point eqn} can be written as
\begin{equation}
     sC_0 + b +(D - d - M)T^{\star}_i + \frac{M}{N} P^{\star} = 0,
\end{equation}
and so
\begin{equation}
     T^{\star}_i =  T^{\star} = \frac{\frac{M}{N}P^\star + b + sC_0}{M + d -D}.
\end{equation}
As $P^\star = N T^\star$, we finally find
\begin{equation}
      T^{\star} = \frac{b + sC_0}{d - D},
\end{equation}
to be the total population size per patch. At long times, when $T_i \rightarrow T^\star$, and assuming $d > D$,  the effective growth rate experienced by each species is 
\begin{equation}
    r_{k,i}^\star  \approx - \frac{bD + dsC_0}{b + sC_0}.
\end{equation} 
As all of the parameters in the above equation are positive, $r_{k,i}^{\star} \leq 0$. This is in agreement with prior work modelling neutral competition between immune cell populations \cite{gaimann2020early}.

\subsection{Metapopulation model}
Consider a single species migrating between $N$ identically coupled spatial sites with independent fluctuating environments. The population size per patch obeys the following dynamics 
\begin{align}
\label{pop eqn}
    \frac{dC_i}{dt} &= \left[r + \eta_i(t)\right]C_i + \frac{M}{N}\sum_{j \neq i}^{N-1} \left(C_j - C_i\right) \nonumber \\
    &= \left[r + \eta_i(t)\right]C_i + M\left(\bar{C} - C_i\right),
\end{align}
where $\bar{C}(t) = \frac{1}{N}\sum_{i=1}^N C_i(t)$ is the average population size over all patches. We change coordinates to the population size of a patch relative to the spatial average population $ w_i = C_i / \bar{C}$ which evolves via a replicator-like equation
\begin{align}
\label{eqn for w}
    \frac{dw_i}{dt} 
    &= \left[ r + \eta_i(t) \right] w_i + M\left(1 - w_i\right) - w_i\left(\frac{1}{N}\sum_{j=1}^N r w_j +\frac{1}{N}\sum_{j=1}^N w_j(t) \eta_j(t)\right) \nonumber \\
    &= M\left(1 - w_i\right) + w_i\left[\eta_i(t) - \zeta(t)\right],
\end{align}
where we have used the fact that by definition $\frac{1}{N} \sum_{i=1}^N w_i = 1$ and defined 
\begin{equation}
    \zeta(t) = \frac{1}{N}\sum_{i=1}^N w_i(t) \eta_i(t).
\end{equation}
To convert the multiplicative noise in Eq.~\eqref{eqn for w} to additive noise, we can shift coordinates to $y_i = \ln w_i$, to obtain 
\begin{equation}
\label{eqn for y}
     \frac{dy_i}{dt}  = M(e^{-y_i} - 1)+\eta_i(t) - \zeta(t)
\end{equation}
In the limit $N \rightarrow \infty$ and with the $\eta_i$'s as uncorrelated white-noises with intensity $2D$, we may write
\begin{equation}
    \frac{1}{N}\sum_{j=1}^N w_j \eta_j(t) dt = \frac{1}{N}\sum_{j=1}^N \sqrt{2D} w_j dB_j + \left(\frac{D}{N}\sum_{j=1}^N w_j - \mathcal{O}\left(\frac{1}{N^2}\right) \right) dt \approx \left<\sqrt{2D} w_j dB_j\right>  + D dt = D dt,
\end{equation}
where we have shifted to the It\^o form and assumed that the system is self averaging so that an average over patches may be replaced by a average over realisations of the noise, denoted by $\left<\right>$. 
If instead the $\eta_i$'s have some degree of temporal correlation, we take $\zeta(t) \rightarrow \zeta$ which must be found self consistently by enforcing $\left<w_i\right>  = 1$.

\section{Coloured noise in the limit of $N\rightarrow \infty$}

\subsection{Steady-state density for $y$}

Taking $\zeta(t) \rightarrow \zeta$ and applying the unified coloured noise approximation to Eq.~\eqref{eqn for y} and dropping the subscript gives
\begin{equation}
\label{ucna eqn for y}
     \frac{dy}{dt}  = \frac{M(e^{-y}-1)-\zeta}{1 + \frac{M}{\lambda} e^{-y}} + \frac{\sqrt{2D}}{1 + \frac{M}{\lambda} e^{-y}} \xi(t).
\end{equation}
We convert this into a Fokker–Planck equation and solve for steady state, finding
\begin{equation}
\label{coloured noise steady state y mf}
    \rho_{\mathrm{mf}}(y) = \frac{1}{Z}  \left(1 + \frac{M}{\lambda}e^{-y}\right)e^{-\frac{M}{D}\left(1 - \frac{1}{\lambda}(M + \zeta)\right) e^{-y}-\frac{M^2}{2\lambda D}e^{-2y} - \frac{1}{D}(M + \zeta)y }, 
\end{equation}
For large $y$, Eq.~\eqref{coloured noise steady state y mf} scales like $\rho_{\mathrm{mf}}(y) \sim e^{-y(M+\zeta)/D}$, and hence $\left<e^{y}\right>$ diverges for large $y$ if $(M + \zeta)/D < 1$. Therefore we require $\zeta > D - M$ for a self consistent solution to exist and the mean-field approximation to hold. We find the normalisation constant to be 
\begin{equation}
    Z = \mathcal{I}_{\nu}(\alpha, \beta) + \frac{M}{\lambda} \mathcal{I}_{\nu+1}(\alpha, \beta),
\end{equation}
where 
\begin{align}
    \mathcal{I}_{\nu}(\alpha, \beta) &= \int_{0}^{\infty}  u^{\nu-1} e^{- \alpha u^2 -\beta u } du = \left(2\alpha\right)^{-\nu/2} \Gamma(\nu) e^{\beta^2/(8\alpha)}\mathcal{D}_{-\nu}\left(\frac{\beta}{\sqrt{2 \alpha}}\right),
\end{align}
where $\mathcal{D}_a(z)$ is a parabolic cylinder function, $\Gamma(z)$ is the Gamma function and we have defined $\alpha = M^2 / (2\lambda D)$, $\beta = (M/D)(1 - (\zeta + M)/\lambda)$ and $\nu = (M + \zeta)/ D$.

\subsection{Steady-state density for $x$}

The steady state density for  $x_{ij} = y_i - y_j, i \neq j$, may be obtained by using $y_i, y_j \overset{i.i.d}{\sim} \rho_{\mathrm{mf}}(y)$. We find the density for $x$ to be,
\begin{align}
\rho_{\mathrm{mf}}(x) &= \int_{-\infty}^{\infty} \rho_{\mathrm{mf}}(y)\rho_{\mathrm{mf}}(y+x) dy \\
    &= \frac{e^{-\nu x}}{Z^2}\left(  \mathcal{I}_{2\nu}(A(x), B(x)) + \frac{M}{\lambda} \left(1 + e^{-x}\right)\mathcal{I}_{2\nu+1}(A(x), B(x)) + \left(\left(\frac{M}{\lambda}\right)^2 + e^{-x}\right) \mathcal{I}_{2\nu+2}(A(x), B(x))\right),
\end{align}
where $A(x) = \alpha \left(1 + e^{-2x}\right)$ and $B(x) = \beta \left(1 + e^{-x}\right)$. The tails of this density scale like $\rho_{\mathrm{mf}}(x) \sim e^{-|x|(M+\zeta)/D }$.

\subsection{Solving for $\zeta$}

$\zeta$ is solved for self consistently by requiring $\left<w\right> = \left<e^y\right> = 1$, which is equivalent to 
\begin{equation}
    \label{self consistency}
    G(\nu, \alpha,\beta) = \mathcal{I}_{\nu-1}(\alpha, \beta)+ \left(\frac{M}{\lambda} - 1\right)\mathcal{I}_{\nu}(\alpha, \beta) - \frac{M}{\lambda}\mathcal{I}_{\nu+1}(\alpha, \beta) = 0.
\end{equation}
From the scaling of the tail of Eq.~\eqref{coloured noise steady state y mf} we require  $\zeta > D - M$, as $\zeta = D \nu -M$ this condition is equivalent to $\nu > 1$.

\subsubsection{Existence of a solution for finite parameters}

For finite $D >0, M>0, \lambda >0$ we can consider $G(\nu, \alpha, \beta)$ as $\nu \rightarrow 1^+$ and $\nu \rightarrow \infty$. First, for $\nu\rightarrow 1^+$, we have $\beta \rightarrow \frac{M}{D} - \frac{M}{\lambda} = \beta_1$. We evaluate each term of Eq.~\eqref{self consistency} in turn
\begin{align}
   \lim_{\nu \rightarrow1^+} I_{\nu}(\alpha, \beta(\nu))  &= \int_{0}^{\infty} e^{-\alpha u^2 - \beta_1 u} du \rightarrow \text{finite} \nonumber \\
      \lim_{\nu \rightarrow 1^+} I_{\nu+1}(\alpha, \beta(\nu))  &= \int_{0}^{\infty} u  e^{-\alpha u^2 - \beta_1 u} du \rightarrow \text{finite} \nonumber \\
      \lim_{\nu \rightarrow1^+} I_{\nu-1}(\alpha, \beta(\nu))  &= \int_{0}^{\infty} \frac{1}{u}  e^{-\alpha u^2 - \beta_1 u} du \rightarrow +\infty \nonumber
\end{align}
Therefore $\lim_{\nu\rightarrow 1^+} G(\nu, \alpha, \beta(\nu)) = +\infty$. We next consider $\nu \rightarrow +\infty$, for which $\beta= \frac{M}{D} - \frac{M}{\lambda} \nu =  \beta_0 - c\nu$. We write 
\begin{align}
  I_{\nu +k}(\alpha, \beta(\nu)) &=  \int_{0}^{\infty} u^{(\nu + k) -1}  e^{-\alpha u^2 - (\beta_0  - c \nu ) u} du \nonumber \\ 
  &=  \nu^{k }\nu^{\nu}\int_{0}^{\infty} t^{\nu + k -1} e^{-\alpha \nu ^2 t^2  - \beta_0 \nu t + c \nu^2 t}dt
\end{align}
where we have changed coordinates to $u = t \nu$. We now make a saddlepoint approximation, defining 
\begin{equation}
    \psi(t) = (\nu + k -1)\ln t - \alpha \nu ^2 t^2  - \beta_0 \nu t + c \nu^2 t,
\end{equation}
which has a critical point defined by 
\begin{align}
    \frac{d\psi}{dt}\Bigg|_{t=t^*} &= \frac{\nu + k -1}{t} - 2\alpha \nu^2 t^{*} - \beta_0 \nu + c\nu^2 = 0  \nonumber \\
    &\implies t^* \approx \frac{c}{2\alpha},
\end{align}
where we have kept terms to leading-order in $\nu$. Evaluating at the saddle and only keeping terms to leading order $\nu$, we have
\begin{equation}
    I_{\nu+k}(\alpha, \beta(\nu)) \approx \nu^{k}\nu^{\nu} e^{\psi(t^{*})} =  \nu^{k}\nu^{\nu} e^{(\nu^2 c^2) / (4\alpha)} = \nu^{k} F(\nu),
\end{equation}
where $F(\nu) = \nu^{\nu} e^{(\nu^2 c^2) / (4\alpha)}$.
Therefore we may write $G(\nu, \alpha,\beta)$ as
\begin{align}
     G(\nu, \alpha,\beta) &\approx \frac{1}{\nu} F(\nu) + \left(\frac{M}{\lambda}  - 1\right) F(\nu) - \frac{M}{\lambda} \nu F(\nu) \nonumber \\
     &= F(\nu)\left(\frac{1}{\nu} + \frac{M}{\lambda}  - 1 - \frac{M}{\lambda} \nu \right) \nonumber \\
     &\approx  -\frac{M}{\lambda} \nu F(\nu)
\end{align}
We now take the limit $\nu \rightarrow +\infty$, writing 
\begin{align}
  -\nu F(\nu) \;\xrightarrow[\nu\to+\infty]{}\; -\infty
\end{align}
Therefore $\lim_{\nu\rightarrow +\infty} G(\nu, \alpha, \beta(\nu)) = -\infty$. As $G(\nu, \alpha, \beta(\nu))$ is continuous on the interval, at least one solution for $G(\nu, \alpha, \beta(\nu)) = 0$ always exists for finite $M, D$ and $\lambda$.

\subsubsection{The quenched limit - $\lambda \rightarrow 0$}
We now take the limit $\lambda \rightarrow 0$,   $\lambda \ll M$, of Eq.~\eqref{coloured noise steady state y mf}, writing
\begin{align}
    \ln \rho_{\mathrm{mf}}(y)  &\sim {-\frac{M}{D}\left(1 - \frac{1}{\lambda}(M + \zeta)\right) e^{-y}-\frac{M^2}{2\lambda D}e^{-2y} - \frac{1}{D}(M + \zeta)y } + \ln \left(1 + \frac{M}{\lambda}e^{-y}\right) \nonumber \\
    &= -\frac{1}{\lambda}\left(\frac{M^2}{2D}e^{-2y} -  \frac{M}{D}(M+\zeta)e^{-y}\right) +\mathcal{O}\left(\ln\left(\frac{1}{\lambda}\right)\right) + \mathcal{O}(1)
\end{align}
In the limit $\lambda\rightarrow 0$, $\rho_{\mathrm{mf}}(y) \rightarrow \delta(y -y^*)$, where $y^*$ is the critical point of 
\begin{equation}
    S(y) = \frac{M^2}{2D}e^{-2y} - \frac{M}{D}(M+\zeta)e^{-y}.
\end{equation}
Therefore
\begin{align}
    \frac{dS}{dy}\Bigg|_{y=y^*} &= -\frac{M^2}{D}e^{-2y^*} + \frac{M}{D}(M + \zeta) e^{-y^*} = 0  \nonumber \\
    &\implies y^* = \ln\left(\frac{M}{M+\zeta}\right).
\end{align}
Enforcing the self consistency condition $\left<e^{y}\right> = 1$, we find that 
\begin{align}
    \lim_{\lambda \rightarrow 0}\left<e^{y}\right> &= \int_{-\infty}^{\infty} \delta(y - y^*) e^{y} dy  = e^{y^*} = \frac{M}{M+\zeta} = 1 \nonumber \\
    &\implies  \zeta = 0
\end{align}
Therefore, in the limit of persistent environmental fluctuations, the condition $\zeta > D -M$, becomes $D < M$ with $\left<w\right>$ diverging for $D \geq M$ and the mean-field approximation failing. If we take $D \lambda = \sigma^2 / (2 \ln N)$, and keep $2\lambda \ln N = M$ as $N \rightarrow \infty$ and $\lambda \rightarrow 0$, we find $\left<w\right>$ diverges for $\sigma^2 \geq M^2$. This matches the condition found for the case of heterogenous random growth rates in~\cite{bernard2025mean}. 

\subsubsection{Weak-noise approximation - $D < M$}

We now study the behaviour of $\zeta$ for $D/M < 1$. We rescale all model parameters with respect to $M$, letting $D =\tilde{D}M$, $\tilde{\lambda} = \lambda M$ and $\tilde{\zeta} = \zeta M$. The condition for a non-diverging mean now becomes $\tilde{\zeta} >\tilde{D} - 1$. Defining $t = e^{-y}$, the self consistency condition may be written as
\begin{equation}
\frac{I_2}{I_1} = \frac{\int_{0}^{\infty} g_{2}(t) \, e^{\Phi(t)/\tilde{D}} \, dt}{\int_{0}^{\infty} g_{1}(t)\, e^{\Phi(t)/\tilde{D}} \, dt} = 1
\end{equation}
where 
\begin{equation}
    g_{\ell}(t) = t^{-\ell} \left(1 + \frac{1}{\tilde{\lambda}}t\right),
\end{equation}
and 
\begin{equation}
    \Phi(t) = (1+\tilde{\zeta})\ln t - \left(1 - \frac{1 + \tilde{\zeta}}{\tilde{\lambda}}\right)t - \frac{1}{2\tilde{\lambda}}t^2.
\end{equation}
For $\tilde{D} \ll 1$, the integrals are dominated by their value around $t^{*}$, the critical point of $\Phi(t)$, such that 
\begin{align}
    \frac{d\Phi}{dt}\Bigg|_{t=t^*} &= (1+\tilde{\zeta}) \frac{1}{t^*} - \left(1 - \frac{1 + \tilde{\zeta}}{\tilde{\lambda}}\right) - \frac{1}{\tilde{\lambda}}t^{*} = 0, \\
    &\implies t^{*} = 1 + \tilde{\zeta}.
\end{align}
Employing the saddlepoint approximation and keeping terms to $\mathcal{O}(\tilde{D})$, each integral may be written as
\begin{equation}
    I_{\ell} \approx \sqrt{\frac{2\pi \tilde{D}}{-\Phi''(t^*)}} e^{\Phi(t^*)/\tilde{D}}\left[g_{\ell}(t^*) + \tilde{D}\left( \frac{g''_{\ell}(t^*)}{-2\Phi''(t^*)} + \frac{g'_{\ell}(t^*)\Phi'''(t^*)}{2(-\Phi''(t^*))^2} + \frac{g_{\ell}(t^*)\Phi''''(t^*)}{8(\Phi''(t^*))^2} + \frac{5g_{\ell}(t^*)(\Phi'''(t^*))^2}{24(-\Phi''(t^*))^3}  \right) + \mathcal{O}\left(\tilde{D}^2\right)\right],
\end{equation}
where prime denotes differentiation with respect to $t$. Computing $I_2/I_1$, the prefactor terms cancel and keeping only terms of $\mathcal{O}(\tilde{D})$, we find
\begin{equation}
    \frac{I_1}{I_2} \approx \frac{1}{1 + \tilde{\zeta}} +\tilde{ D }\left(\frac{\tilde{\lambda}}{(1+\tilde{\zeta})^2(1 + \tilde{\lambda} + \tilde{\zeta})}\right) + \mathcal{O}\left(\tilde{D}^2\right) = 1 .
\end{equation}
This equation has three roots, which to $\mathcal{O}(D)$ are
\begin{equation}
\label{zeta approx}
\tilde{\zeta}_{(1)} = \tilde{D} \left(\frac{\tilde{\lambda}}{\tilde{\lambda} + 1}\right), \quad \tilde{\zeta}_{(2)} = \frac{\tilde{D} - (\tilde{\lambda}+1)^2}{\tilde{\lambda} + 1}, \quad \tilde{\zeta}_{(3)} = -(\tilde{D} + 1).
\end{equation}
For $\tilde{D} < 1$, only the first root is greater than $\tilde{D}-1$. Converting back to dimensional parameters, this root is 
\begin{equation}
    \zeta_{(1)} = D\left(\frac{\lambda}{\lambda + M}\right).
\end{equation}

\subsection{The long-term population growth rate}

The long-term logarithmic growth rate of the population is 
\begin{equation}
    \left<\frac{d\ln \bar{C}}{dt}\right> = r +\left<\frac{1}{N}\sum_{i=1}^N w_i(t) \eta_i(t) \right> = r + \left<\zeta(t)\right> \rightarrow r + \zeta.
\end{equation}
As migration is symmetric, the long-term growth rate may alternatively be computed by considering the growth of a single patch, for $N=2$ we have
\begin{equation}
\label{two patch growth rate}
    \left<\zeta\right> = \left<\frac{d\ln C_1}{dt}\right> - r = \frac{M}{2} \left( \left<e^{y_2 - y_1}\right> - 1 \right)  = \frac{M}{2} \left( \left<e^{-x}\right>_{\rho(x)} - 1 \right),
\end{equation}
where we have used the fact that $\rho(x, t)$ approaches a steady state. For white noise, the expectation may be solved for analytically with \cite{agranov2021extinctions, bouchaud2015growth}
\begin{equation}
    \left<\zeta\right> = \frac{M}{2} \left(\frac{\mathcal{K}\left(-1, \frac{M}{2D}\right)}{\mathcal{K}\left(0, \frac{M}{2D}\right)} - 1\right),
\end{equation}
where $\mathcal{K}(\nu, z)$ is a modified Bessel function of the second kind. For coloured noise at $N=2$ we evaluate Eq.~\eqref{two patch growth rate} numerically. 

\begin{figure}[h!]
\includegraphics[width=0.4\columnwidth]{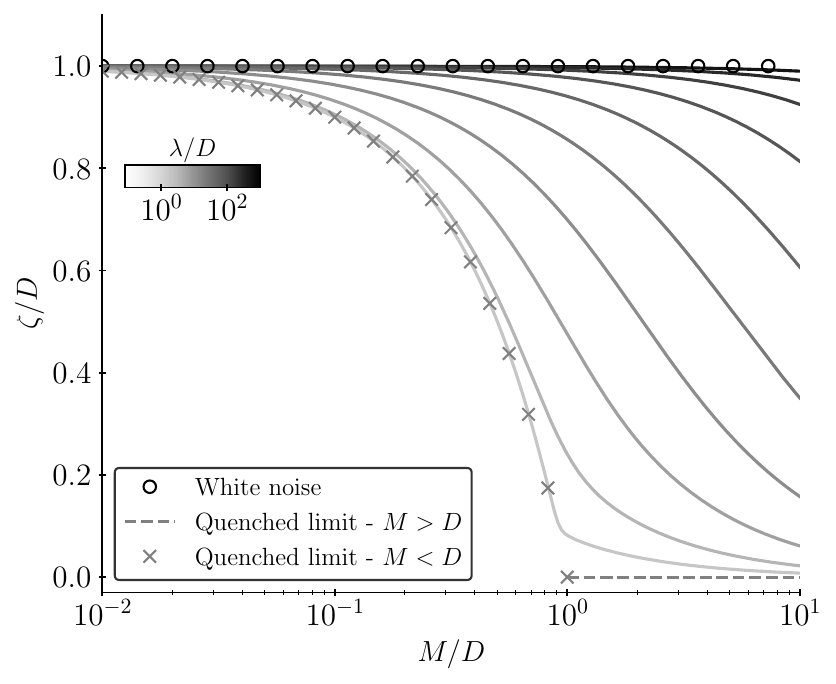}
\caption{Logarithmic growth rates for a species migrating between time varying environments in the limit of $N \rightarrow \infty$. Each curve shows the numerically obtained growth rates for environmental fluctuations with a correlation timescale of $1/\lambda$. Circles are the analytic solution found in the case of temporally uncorrelated environmental fluctuations. The dashed line is the analytically obtained growth rate in the limit $\lambda/D \rightarrow 0$, which is valid for $D < M$, beyond which the mean-field approximation breaks down. The crosses are the predictions of the growth rate from \cite{bernard2025mean} which studies the case of quenched random growth rates. We make this comparison was made by taking  $D \lambda = \sigma^2 / (2 \ln N)$, and keeping $2\lambda \ln N = M$ as $N \rightarrow \infty$ and $\lambda \rightarrow 0$ such that $\sigma^2 \approx MD$. The authors find that for $M > \sigma$, the long term growth rate of the system is $\zeta = 0$, while for   $M < \sigma$ they find $\zeta = \sigma-M$. All rates have been rescaled by the intensity of the environmental fluctuations, $D$.}
\label{fig growth rate mf coloured}
\end{figure}

\section{Bimodality emerging from heterogeneous random growth rates for $N=2$}

In the absence of environmental fluctuations ($D \rightarrow 0$), the log-ratio coordinate for two patches with non-equal basal growth rates obeys the following differential equation
\begin{equation}
    \frac{dx}{dt} = s -M\sinh x,
\end{equation}
where $s = r_1 -r_2$. At long times $x$ approaches an attractive fixed point, $x^\star$, given by
\begin{equation}
    x^\star = \sinh^{-1}\left(\frac{s}{M}\right).
\end{equation}
Consider an ensemble of systems for which each $r_i$ is drawn from a normal distribution of mean $\mu_r$ and variance $\sigma_r^2$ such that $s \sim \mathcal{N}(0, 2\sigma_r^2)$. The distribution of fixed points over this ensemble, $\rho(x^\star)$, is then given by 
\begin{align}
    \rho(x^\star) &= \rho\left(s(x^\star)\right) \left| \frac{ds}{dx^\star}\right|  \nonumber \\ 
    &= \frac{M}{\sqrt{4\pi \sigma_r^2}}  e^{-\frac{M^2}{4\sigma_r^2} \sinh^2x^\star} \cosh x^\star. 
\end{align}
The extrema of $\rho(x^\star)$ may be identified as the extrema of  
\begin{equation}
    - \log \rho(x^\star) \propto \frac{M^2}{4\sigma_r^2} \sinh^2x^\star - \ln \left(\cosh x^\star\right).
\end{equation}
Which has turning points at $x^\star_0 = 0$ and $x^\star_{\pm} = \pm\cosh^{-1} \left(\sqrt{2} \frac{\sigma_r}{M}\right)$. These last two extrema are maxima of $\rho(x^\star)$ and only defined for $2\sigma_r^2/M^2 \geq 1$. Therefore, $\rho(x^\star)$ is bimodal for $\sigma_r^2 \geq M^2/2$ (including the case of degenerate roots) in agreement with the quenched, $\lambda \rightarrow 0$, limit of coloured noise. 

\section{Simulation scheme}

Numerical simulations of the stochastic differential equations presented in the main text were performed using the Euler-Maruyama method, with Ornstein-Uhlenbeck processes being simulated using exact discretisation. For $N=2$ driven by white noise, the update was
\begin{equation}
   x(t+\Delta t) = x(t) - M\sinh x(t) \Delta t + \sqrt{4D\Delta t} \xi,
\end{equation}
with $\xi \sim \mathcal{N}(0, 1)$. For $N\rightarrow \infty$, Eq.~\eqref{eqn for y} was simulated with $\zeta = D$ such that the update was
\begin{equation}
   y(t+\Delta t) = y(t) + \left(M e^{-y(t)} -M - D\right) \Delta t + \sqrt{2D\Delta t} \xi.
\end{equation}
Trajectories of $x$ were computed from $y_{i} - y_j$, where the index denotes independent simulation runs. For $N =2$ driven by coloured noise
\begin{align}
   x(t+\Delta t) &= x(t) + \left(-M\sinh x(t) + \eta_x(t) \right)\Delta t, \nonumber \\ \nonumber \\
   \eta_x(t+\Delta t) &= e^{-\lambda \Delta t} \eta_x(t) + \sqrt{2D\lambda \left(1 - e^{-2\lambda \Delta t}\right)} \xi
\end{align}
For $N \rightarrow \infty$ driven by coloured noise, we used the following update
\begin{align}
\zeta(t) &= \frac{1}{n_{\text{runs}}}\sum_{i=1}^{n_{\text{runs}}} e^{y_i(t)} \eta_i(t) \nonumber \\ \nonumber \\
   y_i(t+\Delta t) &= y_i(t) + \left(M e^{-y_i(t)} -M - \zeta(t) + \eta_i(t)\right) \Delta t , \nonumber \\ \nonumber \\
   \eta_i(t+\Delta t) &= e^{-\lambda \Delta t} \eta_i(t) + \sqrt{D\lambda \left(1 - e^{-2\lambda \Delta t}\right)} \xi_i
\end{align}
where $i$ spans simulation runs updated in parallel.

\section{Additional figures}

\begin{figure}[h!]
\includegraphics[width=\columnwidth]{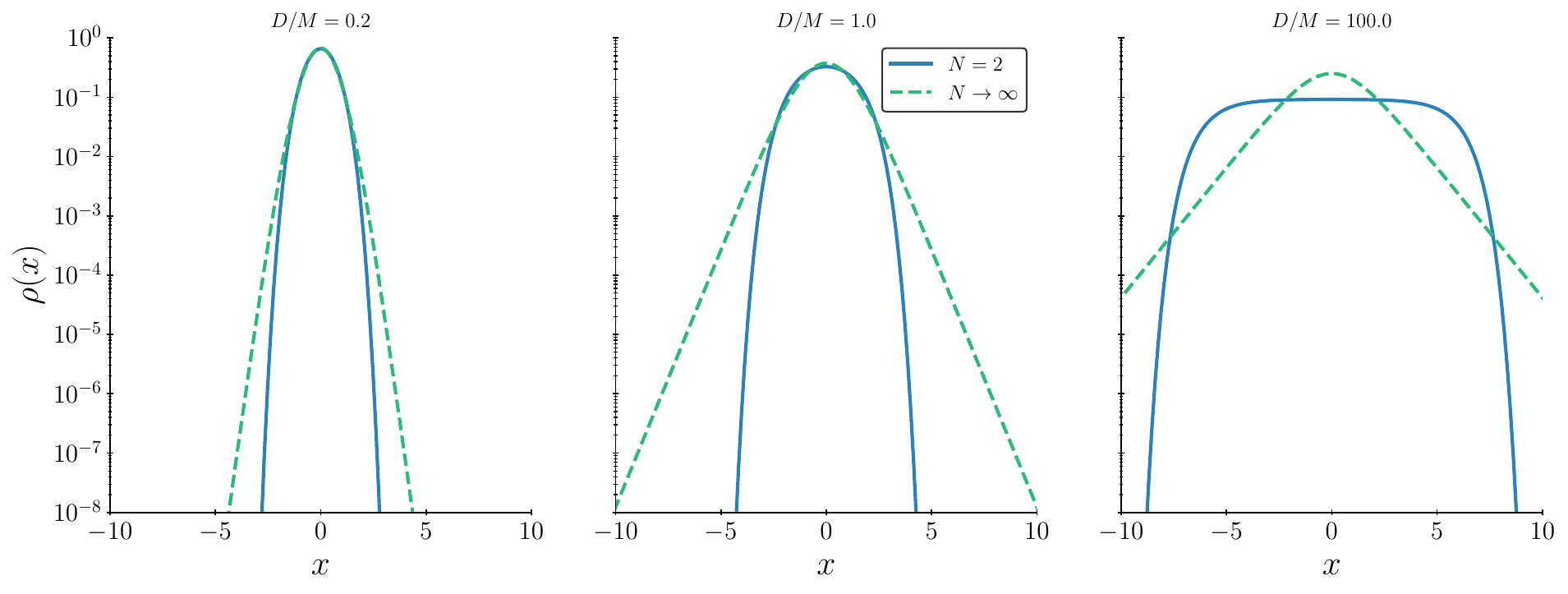}
\caption{Analytical densities for the steady-state distribution of two-point inequality in the limits of two patches and an infinite number of patches for a range of noise intensity, $D$, to migration rate, $M$, ratios. Plots highlight the differences in tail behaviour, with exponentially decaying tails in log space for $N=2$ and linear tails for $N \rightarrow \infty$. We anticipate that any system with finite $N$ should feature exponentially decaying tails in log space for large $|x|$.}
\label{fig tail scaling}
\end{figure}

\begin{figure}[h!]
\includegraphics[width=\columnwidth]{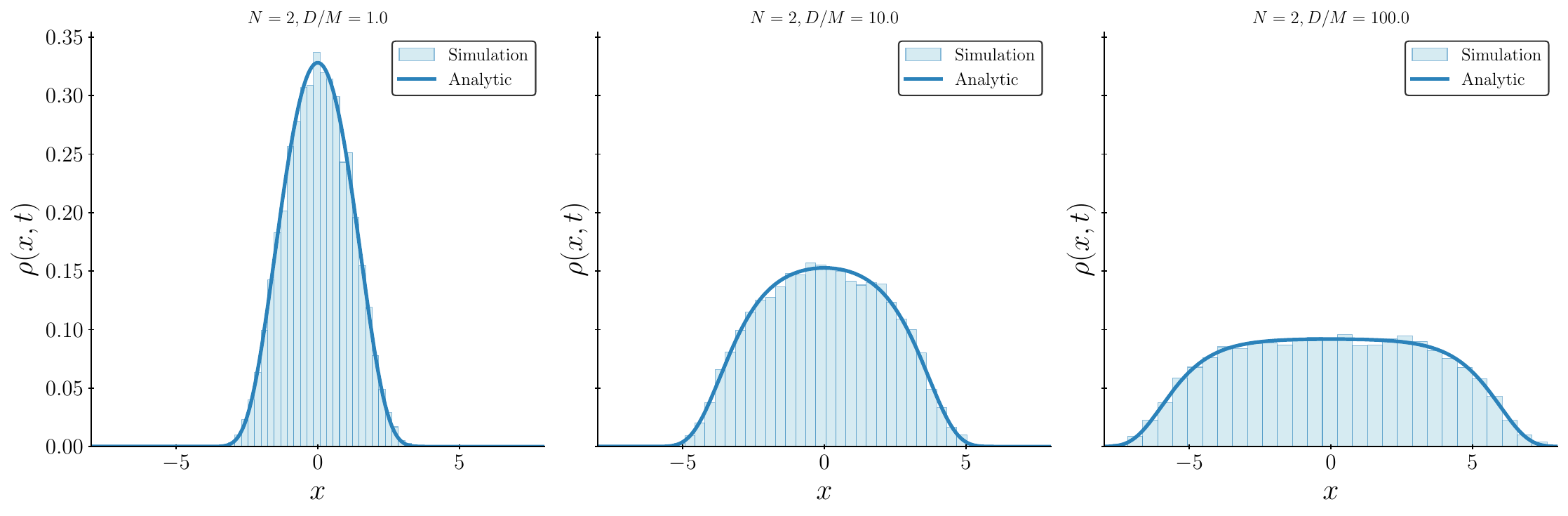}
\caption{Histograms of trajectories of $x$ for a two patch system driven by white noise. Results obtained at $Mt = 20 $ over $2 \times 10^4$ simulation runs. The analytic steady state densities are overlaid.}
\label{fig numerics vs sim white noise two patch}
\end{figure}
\begin{figure}[h!]
\includegraphics[width=\columnwidth]{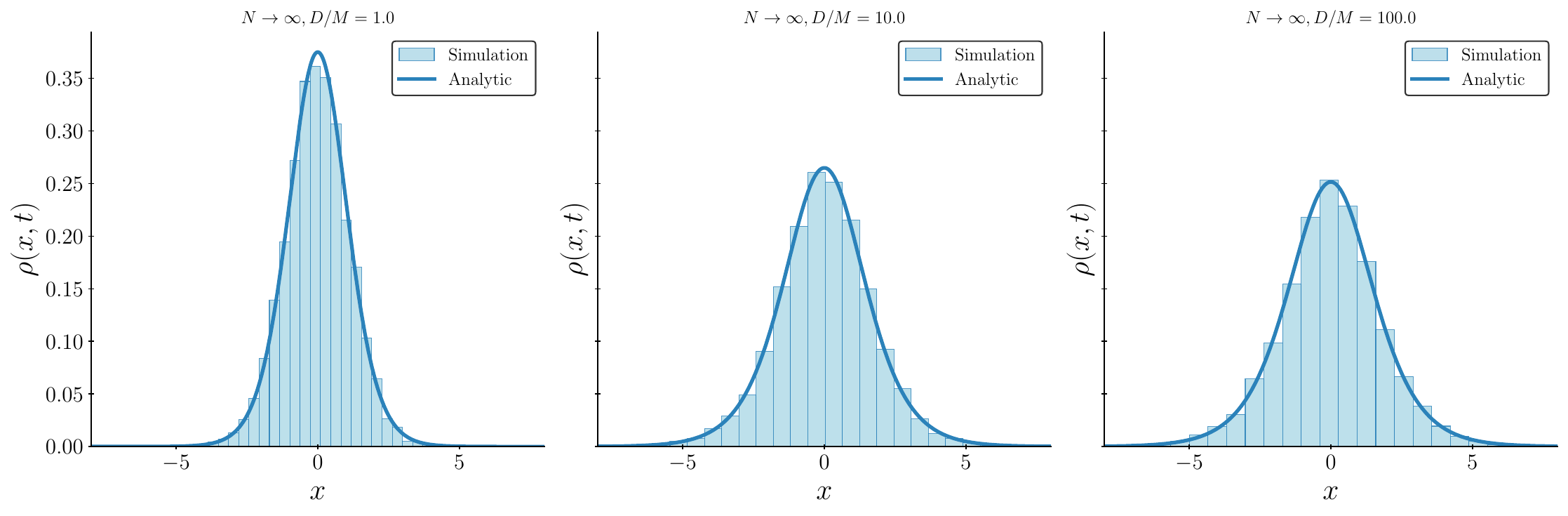}
\caption{Histograms of trajectories of $x$ for an infinite patch system driven by white noise. Simulations were performed using stochastic differential equations obtained via a mean field approximation. Results obtained at $Mt = 20 $ over $2 \times 10^4$ simulation runs. The analytic steady state densities are overlaid.}
\label{fig numerics vs sim white noise mf}
\end{figure}

\begin{figure}[h!]
\includegraphics[width=\columnwidth]{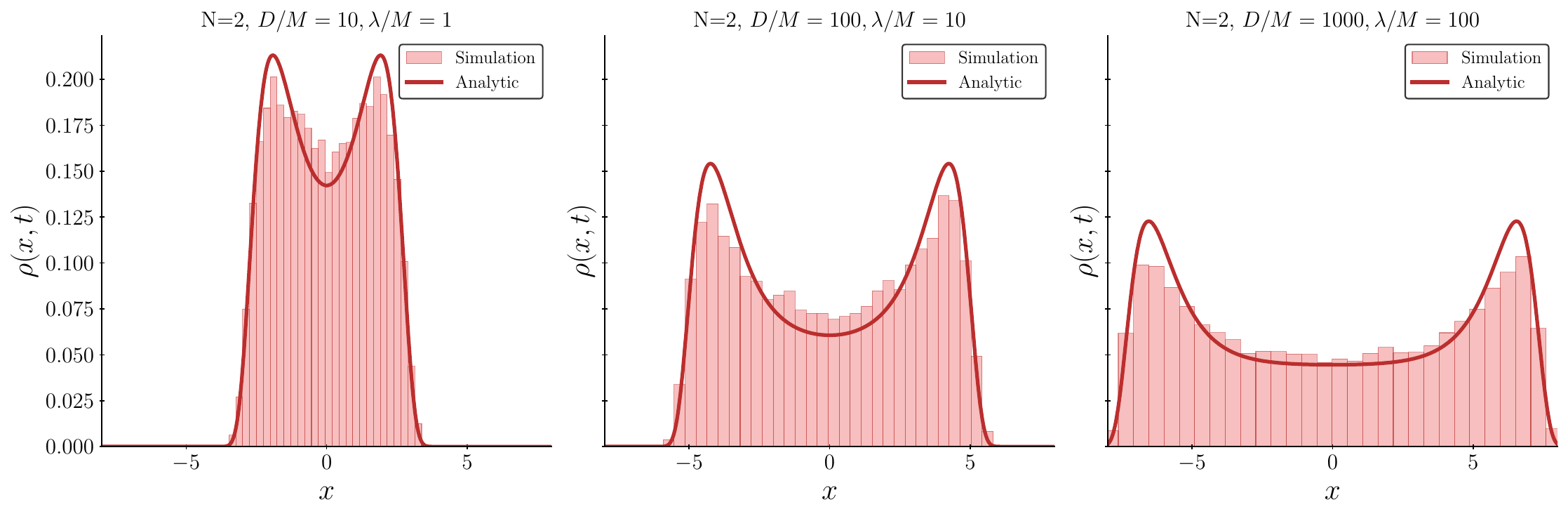}
\caption{Histograms of trajectories of $x$ for a two patch system driven by coloured noise. Results obtained at $Mt = 20 $ over $2 \times 10^4$ simulation runs. The analytic steady state densities are overlaid.}
\label{fig numerics vs sim coloured noise two patch}
\end{figure}
\begin{figure}[h!]
\includegraphics[width=\columnwidth]{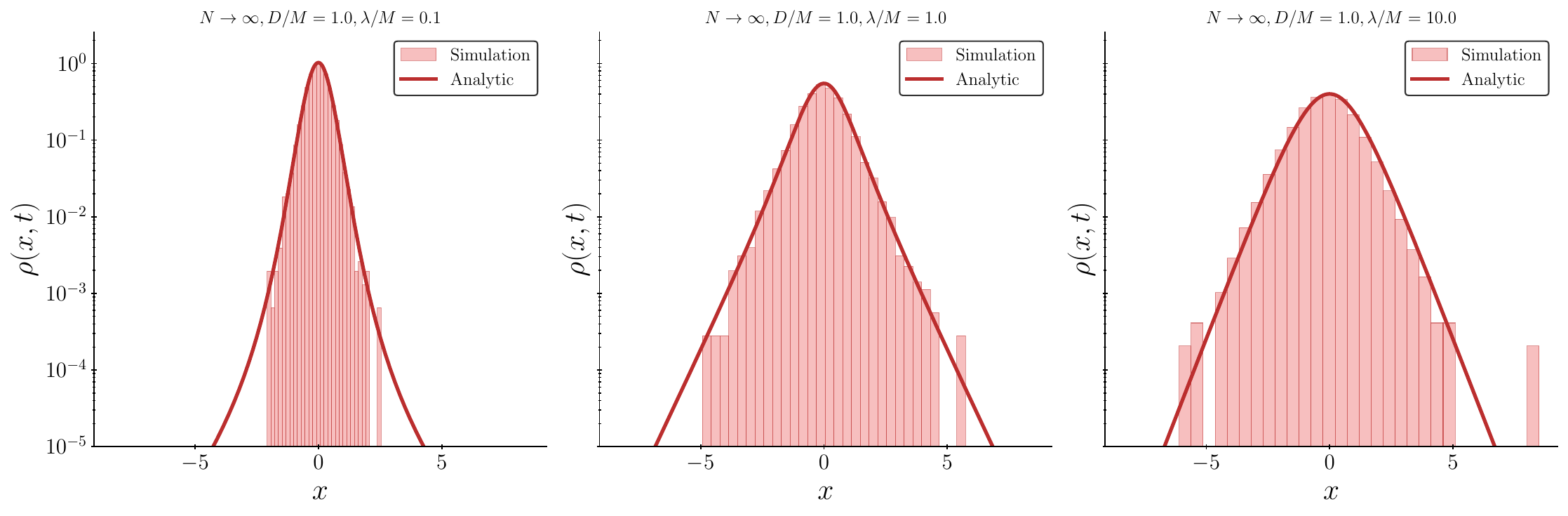}
\caption{Histograms of trajectories of $x$ for an infinite patch system driven by coloured noise. Simulations were performed using stochastic differential equations obtained via a mean field approximation. Results obtained at $Mt = 20 $ over $2 \times 10^4$ simulation runs. The analytic steady state densities are overlaid.}
\label{fig numerics vs sim coloured noise mf}
\end{figure}

\begin{figure}[h!]
\includegraphics[width=\columnwidth]{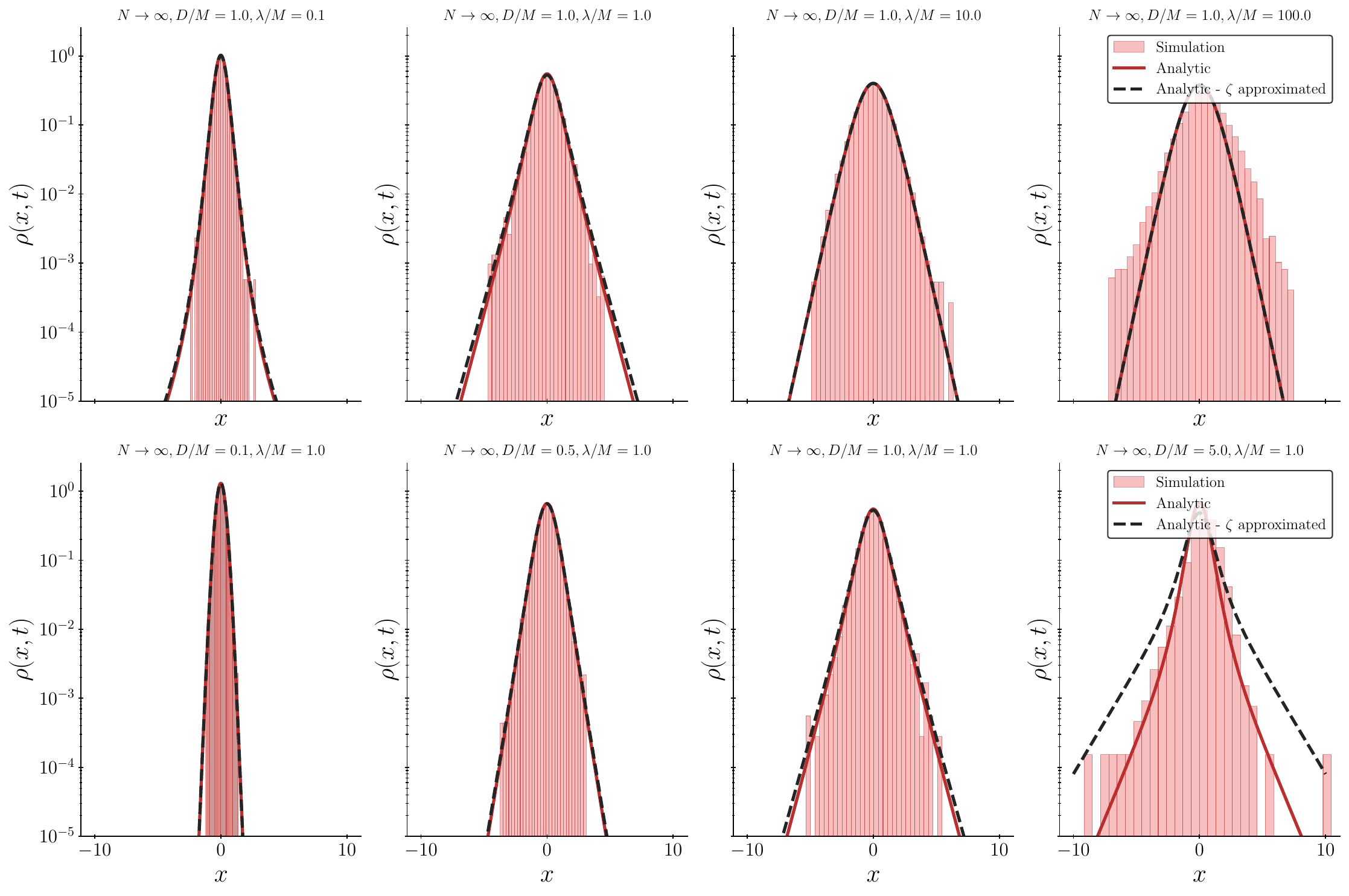}
\caption{Histograms of trajectories of $x$ for an infinite patch system driven by coloured noise. Simulations were performed using stochastic differential equations obtained via a mean field approximation. Results obtained at $Mt = 20 $ over $2 \times 10^4$ simulation runs. The analytic steady state densities are overlaid in addition to the analytic densities using an approximation obtained using a saddlepoint approximation for $\zeta$ for weak noise ($D < M$).}
\label{fig approximate zeta}
\end{figure}

\clearpage
\putbib
\end{bibunit}

\end{document}